\documentstyle[prl,twocolumn,aps]{revtex}
\begin{document}
\input psfig
\draft

\title{Cotunneling at resonance for the single-electron transistor}

\author{J\"urgen K\"onig, Herbert Schoeller and Gerd Sch\"on}

\address{
Institut f\"ur Theoretische Festk\"orperphysik, Universit\"at
Karlsruhe, 76128 Karlsruhe, Germany}

\date{\today}

\maketitle

\begin{abstract}

We study electron transport through a small metallic island in the
perturbative regime. Using a recently developed diagrammatic technique, we 
calculate the occupation of the island as well as the conductance through the 
transistor in forth order in the tunneling matrix elements, a process
referred to as cotunneling. Our formulation does not require the introduction 
of a cut-off. At resonance we find significant modifications of previous 
theories and good agreement with recent experiments.

\end{abstract}

Electron transport through small metallic islands is strongly influenced 
by the charging energy associated with low capacitance of the junctions 
\cite{Ave-Lik,Gra-Dev,Sch-Uebersicht}.
A variety of single-electron effects, including Coulomb blockade phenomena and 
gate-voltage dependent oscillations of the conductance, have been observed. 
Usually, they are described within the ``orthodox theory''\cite{Ave-Lik} which 
treats tunneling in lowest order perturbation theory (golden rule) and 
corresponds to the classical picture of incoherent tunneling processes 
(sequential tunneling). 
As a necessary condition one needs weak tunneling, i.e., the conductance of 
the barriers has to be low
\begin{equation}
	\alpha_0\equiv h/(4\pi^2 e^2 R_{\rm T}) \ll 1 . 
\end{equation}
Despite the success of this straightforward approach, it was found
experimentally and theoretically that there are several regimes where higher 
order tunneling processes have to be taken into account. 

First, in the Coulomb blockade regime, sequential tunneling is exponentially
suppressed. The leading contribution to the current is a second-order process 
in which electrons tunnel via a virtual state of the island.
Averin and Nazarov  \cite{Ave-Naz} evaluated  the transition rate of this 
``inelastic cotunneling'' process  at zero temperature.
At finite temperature, divergences arise, but the authors of 
Ref.~\cite{Ave-Naz} provide an approximation which is valid far away from the 
resonances. 
They assumed that some regularization procedure will overcome the divergences. 
Their results were confirmed experimentally \cite{Exp-Cot}.

Second, it was found recently \cite{SS,KSS1} that even at resonance,
where sequential tunneling is not suppressed, higher order processes
are important and can lead to a significant change of the
conductance. Similar effects were discussed for the average charge of the
single-electron box in equilibrium 
\cite{Laf,Mat,Gol,Fal-Schoen-Zim,Gra}.  
A diagrammatic real-time technique was developed for 
metallic islands \cite{SS,KSS1} as well as for quantum dots
\cite{KSS2,KSSS} in order to give a systematic description of the
various tunneling processes. The effects from quantum
fluctuations were shown to 
become observable either for strong tunneling $\alpha_0\sim 1$ or at
low enough temperatures $\alpha_0 \ln{E_{\rm C}/T} \sim 1$, where $E_{\rm C}$
denotes the charging energy. The predicted broadening of the
conductance peak as well as the reduction of its height was confirmed
in the experiments of Joyez et al. \cite{esteve}
in the strong tunneling regime. 
Within the theory, only processes where the two classically occupied
charge states are involved (even virtually) were included.
Therefore, it was necessary to introduce a band-width cut-off 
$\sim E_{\rm C}$, which prohibits a comparison with experiment without 
fitting parameters.
However, at low temperatures, a quantitative fit between theory and experiment 
is possible provided a renormalized value for the charging energy is used as 
it was determined in the experiment.

In this Letter, we use the same diagrammatic technique to obtain the total
current in second order in $\alpha_0$ including {\it all} relevant processes
such that no cut-off remains.
All terms are regularized in a natural way.
This is important for a comparison with experiments since only system 
parameters enter the result, while a cut-off or some arbitrary regularization
scheme would be an undetermined ingredient in the theory.
At resonance we obtain new contributions compared to the existing
theory of electron cotunneling. They emerge from a change of the
occupation probabilities and a renormalization of the charge excitation energy.
For realistic parameters $T/E_{\rm C}\sim 0.05$  
and $\alpha_0^{\rm L}=\alpha_0^{\rm R} \sim 0.02$ the
corrections are of order $20\%$. We compare with recent experiments
\cite{esteve} and find reasonable agreement 
without fitting any parameter.

The single-electron transistor is modeled by the standard tunneling Hamiltonian
\begin{equation}
	H=H_{\rm L}+H_{\rm R}+H_{\rm I}+H_{\rm ch}+H_{\rm T}=H_0+H_{\rm T}.
\end{equation}
Here $H_{\rm r}=\sum_{kn}\epsilon^{\rm r}_{kn}a^\dagger_{{\rm r}kn}
a_{{\rm r}kn}$ and $H_{\rm I}=\sum_{qn}\epsilon_{qn} c^\dagger_{qn} c_{qn}$
describe the noninteracting electrons in the two leads r=L,R and on
the island where $n$ is the transverse channel index which includes the
spin. The wave vectors $k$ and $q$ numerate the states of the
electrons for fixed r and $n$. In the following, we consider ``wide''
metallic junctions with $N \gg 1$ transverse channels.
The Coulomb interaction of the electrons on the island is modeled by
$H_{\rm ch}=E_{\rm C}(\hat{n}-n_{\rm x})^2$,
where $E_{\rm C}={e^2\over 2C}$ is the charging energy, 
and $en_{\rm x}=C_{\rm L} V_{\rm L}+C_{\rm R} V_{\rm R}+C_{\rm g} V_{\rm g}$, 
with $C=C_{\rm L}+C_{\rm R}+C_{\rm g}$, describes an external charge,
which depends  
on the voltages and capacitances of the circuit.
The excess particle number operator on the island is given by $\hat{n}$.
The charge transfer processes due to tunneling are described by 
\begin{equation}
	H_{\rm T}=\sum_{\rm r=L,R}\sum_{kqn}(T^{{\rm r}n}_{kq}
	a^\dagger_{{\rm r}kn}c_{qn}e^{-i\hat{\varphi}}+h.c.)
\end{equation}
where $T^{{\rm r}n}_{kq}$ are the tunneling matrix elements and 
$e^{\pm i\hat{\varphi}}$ changes the excess particle number on the island by 
$\pm 1$.
The tunneling matrix elements $T^{{\rm r}n}_{kq}=T^{{\rm r}n}$ are considered 
independent of the states $k$ and $q$.
They are related to the tunneling resistances $R_{\rm T,r}$ of the left and 
right junction via ${1\over R_{\rm T,r}} = {2\pi e^2\over \hbar}
\sum_n N^n_{\rm r}(0) N^n_I(0) |T^{{\rm r}n}|^2$,
where $N^n_{\rm I}(0)$ and $N^n_{\rm r}(0)$ are the density of states of the 
island and the leads.

\begin{figure}
\centerline{\psfig{figure=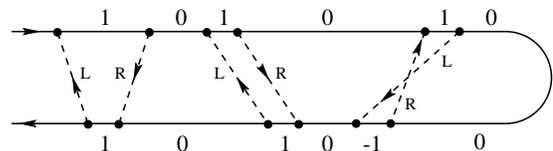,height=1.9cm}}
\caption{A diagram showing contributions to sequential tunneling 
	($\Sigma^{\rm L,-}_{0,1}$ and $\Sigma^{\rm R,+}_{1,0}$) and cotunneling
	($\Sigma^{\rm R,+}_{0,0}$ and $\Sigma^{\rm L,-}_{0,0}$).}
\label{fig1}
\end{figure}

In the following we use the diagrammatic technique developed in 
Ref.~\cite{SS,KSS1}.
The nonequilibrium time evolution of the charge degrees of freedom on the 
island is described by its reduced density matrix, which we obtain in an 
expansion in $H_{\rm T}$.
The reservoirs are assumed to remain in thermal equilibrium (with 
electro-chemical potential $\mu_{\rm r}$) and are traced out by using Wick's 
theorem, such that the Fermion operators are contracted in pairs.
For a large number of transverse channels $N$, only those 
configurations contribute where the operators $a^\dagger_{{\rm r}kn}c_{qn}$
at one time, i.e., from one term $H_{\rm T}$ are contracted with the 
operators $c^\dagger_{qn}a_{{\rm r}kn}$ from {\it one} other term $H_{\rm T}$, 
and not from different ones.
These ``simple loops'' dominate over more complicated configurations with 
more than two times connected by contractions.
Matrix elements of the reduced density operator are visualized in 
Fig.~(\ref{fig1}).
The forward and the backward propagator (Keldysh contour) are coupled by 
``tunneling lines'' (simple loops) associated with the junctions to the 
reservoirs r. 
Each tunneling line with energy $\omega$ represents the rate 
$\alpha^\pm_{\rm r}(\omega)$ if the line is directed backward (forward) with 
respect to the closed time path with
\begin{equation}
	\alpha^\pm_{\rm r}(\omega) = \pm \alpha_0^{\rm r} {\omega-\mu_{\rm r} 
	\over \exp[\pm \beta(\omega-\mu_{\rm r})]-1} \, .
\end{equation}
They are associated with changes of the charge state, as indicated on the 
closed time path.
Finally, we associate with each tunneling vertex at time $t$ a factor 
$\exp{(i\Delta E\,t)}$ where $\Delta E$ is the difference of out- and incoming 
energies.
If the vertex lies on the backward propagator it acquires a factor $-1$. 
We define $\alpha_0=\sum_{\rm r} \alpha_0^{\rm r}$ and 
$\alpha (\omega)=\sum_{\rm r} \alpha_{\rm r} (\omega)$.

The transport properties are affected by the energy difference of adjacent 
charge states
\begin{equation}
	\Delta_n=E_{\rm ch}(n+1)-E_{\rm ch}(n)=E_{\rm C}[1+2(n-n_{\rm x})] \, .
\end{equation}

The formally exact Master equation reads in the stationary case \cite{SS,KSS1}
\begin{equation}
	\dot{p}_n =
	\sum\limits_{n'\neq n}\left[ 
	p_{n'} \Sigma_{n',n} - p_n \Sigma_{n,n'} \right] =0 
\end{equation}
with the probability $p_n$ to be in charge state $n$ and transition rates
$\Sigma_{n,n'}$ to change from state $n$ to $n'$.
In the perturbative regime we write
$p_n=p_n^{(0)}+p_n^{(1)}+p_n^{(2)}+\ldots$ and
$\Sigma_{n,n'}=\Sigma_{n,n'}^{(1)}+\Sigma_{n,n'}^{(2)}+\Sigma_{n,n'}^{(3)}
+\ldots$
where $p_n^{(k)}$ and $\Sigma_{n,n'}^{(k)}$ denotes the term 
$\sim \alpha_0^k$ of the expansion.
The Master equation must hold in each order.
In lowest order (sequential tunneling) it reads 
$p_n^{(0)} \alpha^+(\Delta_n) - p_{n+1}^{(0)} \alpha^-(\Delta_n) = 0. $
At low temperature at most two charge states ($n=0,1$) are important, all 
other states are suppressed exponentially.

Due to higher order processes, however, the occupation is modified and 
also the probability for the other charge states can be nonzero (they are 
algebraically suppressed, but not exponentially). 
The Master equation expanded in second order gives a relation between the rates
in second order $\Sigma_{n,n'}$ (diagrams with two lines) and the occupation
in first order $p_n^{(1)}$ which lead to a deviation in the average 
occupation $\langle n \rangle = \sum_n n p_n= \langle n \rangle^{(0)} +
\langle n \rangle^{(1)}+ \ldots$.

The stationary current 
$I_{\rm r}=-ie\sum_{n,n'}p_n \Sigma_{n,n'}^{\rm r+}$ 
through reservoir $r$ uses the rates $\Sigma_{n,n'}^{\rm r +}$ where the 
rightmost tunneling line corresponds to reservoir r and is an outgoing 
(incoming) one if the rightmost vertex lies on the upper (lower) propagator
(and vice versa for $\Sigma_{n,n'}^{\rm r -}$). 
There are two types of diagrams contributing to the second order correction of 
the current $I^{(2)}$: 
those of the form $p^{(0)}\Sigma^{(2)}$ and the others like 
$p^{(1)}\Sigma^{(1)}$.
The first ones correspond to ``new'', second order processes, and the second 
ones are responsible for a modification of the first order processes due to 
the fact, that the occupation probabilities are changed in higher orders.
The latter have not been considered in previous theories. As we will
see later, both contributions are important. 

In lowest order the average occupation
\begin{equation}
	\langle n \rangle^{(0)} = \alpha^+(\Delta_0) / \alpha (\Delta_0) 
\end{equation}
is only smeared by temperature.
Quantum fluctuations, however, yield
\begin{eqnarray}\label{average charge}
	\langle n \rangle^{(1)}=-{1\over 2E_{\rm C}}{\partial \over \partial 
	n_{\rm x}}
	\left[ p_0^{(0)} \left( {\rm Re} \, \sigma^-(\Delta_{-1}) - 
				{\rm Re} \, \sigma^+(\Delta_0)  \right) \right.
	\nonumber \\ \left.
	     + p_1^{(0)} \left( {\rm Re} \, \sigma^-(\Delta_0) - 
				{\rm Re} \, \sigma^+(\Delta_1)  \right) \right]
\end{eqnarray}
where ${\rm Re}\, \sigma^\pm(\omega) = \sum_{\rm r=L,R} {\rm Re}\, 
\sigma^\pm_{\rm r}(\omega)$ and
\begin{eqnarray}
	{\rm Re} \, \sigma^\pm_{\rm r}(\omega) =
	{\rm Re} \, \int\limits_{-\infty}^{\infty} d\omega'
	{\alpha^\pm_{\rm r} (\omega') \over \omega -\omega' + i0^+} =
	\nonumber \\
  	\alpha_0^{\rm r} 
	\left\{ \pm {\pi \over 2}U -
 	(\omega -\mu_{\rm r}) \left[ \ln {\beta U\over 2 \pi} - 
	{\rm Re} \, \Psi \left(i{\beta\over 2\pi} (\omega -\mu_{\rm r})\right) 
	\right] \right\} \, .
\end{eqnarray}
The integrals $\sigma^\pm(\omega)$ are divergent since the integrand does not 
decay to zero for $|\omega| \rightarrow \infty$.
Therefore, we introduce a Lorentzian cut-off with cut-off parameter $U$.
For the physical quantities like $\langle n \rangle$, however, only 
combination of terms occur where this cut-off drops out, i.e., the divergences 
of different integrals cancel.

In equilibrium, i.e. at $V=0$, the transistor is equivalent to the single
electron box.
A systematic perturbative expansion of the partition function (up to order 
$\alpha_0^2$) was performed by Grabert \cite{Gra}.
The result Eq.~(\ref{average charge}) is identical to Grabert's result 
in order $\alpha_0$, which, at zero temperature, reads 
$\langle  n \rangle^{(1)}=\alpha_0 \ln [(1+2n_{\rm x})/(1-2n_{\rm x})]$.
As a generalization Eq.~(\ref{average charge}) also applies for the
nonequilibrium situation, i.e. $V \neq 0$.

The current $I=I_{\rm L}=-I_{\rm R}$ is in lowest order given by
\begin{eqnarray}
	I^{(1)}(\Delta_0)={4 \pi^2 e\over h}
	{\alpha_{\rm L}(\Delta_0) \alpha_{\rm R}(\Delta_0) \over 
	\alpha(\Delta_0)}
	\left[ f_{\rm L}(\Delta_0) - f_{\rm R}(\Delta_0) \right] \, . 
\end{eqnarray}
The cotunneling contribution can be divided into three parts
$I^{(2)}(\Delta_0)=\sum_{i=1}^3 I^{(2)}_i (\Delta_0)$
with
\begin{eqnarray}\label{current1}
	I^{(2)}_1(\Delta_0)=
	- \int d\omega \, I^{(1)}(\omega) \alpha(\omega)
	\nonumber \\ 
	{\rm Re} \,  \left[ p_0^{(0)} 
	\left( {1 \over \omega - \Delta_0 + i0^+}
	     - {1 \over \omega - \Delta_{-1} + i0^+} \right)^2 
	\right. \nonumber \\ \left. +
	p_1^{(0)} 
	\left( {1 \over \omega - \Delta_1 + i0^+}
	     - {1 \over \omega - \Delta_0 + i0^+} \right)^2  \right] \, ,
\end{eqnarray}
\begin{eqnarray}\label{current2}
	I^{(2)}_2(\Delta_0)= I^{(1)}(\Delta_0)
	\nonumber \\ 
	\int d\omega \, {\rm Re} \, \left[
	\left( {\alpha^-(\omega) \over \omega - \Delta_0 + i0^+}
	     - {\alpha^-(\omega) \over \omega - \Delta_{-1} + i0^+} \right)^2 
	\right. \nonumber \\ \left. +
	\left( {\alpha^+(\omega) \over \omega - \Delta_1 + i0^+}
	     - {\alpha^+(\omega) \over \omega - \Delta_0 + i0^+} \right)^2 
	\right] \, ,
\end{eqnarray}
\begin{eqnarray}\label{current3}
	I^{(2)}_3(\Delta_0)=
	\left[ {\partial \over \partial \Delta_0} I^{(1)}(\Delta_0) \right] \, 
	\nonumber \\
	\int d\omega \, {\rm Re} \, \left[
          {\alpha^-(\omega) \over \omega - \Delta_0 + i0^+}
	 -{\alpha^-(\omega) \over \omega - \Delta_{-1} + i0^+}
	\right. \nonumber \\ \left.
         -{\alpha^+(\omega) \over \omega - \Delta_1 + i0^+} 
	 +{\alpha^+(\omega) \over \omega - \Delta_0 + i0^+} 
        \right] \, .
\end{eqnarray}
The poles at $\omega =\Delta$ are regularized in a natural way (it comes out 
of our theory and is {\it not} added by hand) as Cauchy's principal values 
${\rm Re} {1 \over x+i0^+} = P {1\over x}$ and their derivative 
${\rm Re} {1 \over (x+i0^+)^2} = -{d\over dx} P {1\over x}$.

Deep in the Coulomb blockade regime, we have $p_0^{(0)}=1$, $p_1^{(0)}=0$ and 
$I^{(1)}(\Delta_0)=0$.
Consequently, the first line of $I^{(2)}_1$ is the only contributing one. 
At $T=0$, the integrand is zero at the poles, and we can omit the term 
$+i0^+$. This gives the well-known result of inelastic cotunneling
\cite{Ave-Naz}. At finite temperature, however, the regularization
scheme is needed which is not provided by previous theories within
second order perturbation theory \cite{com3}. Our result is also
well-defined for $T \neq 0$. 

Furthermore, we are able to describe the system at resonance.
In this regime, $I^{(2)}_2$ and $I^{(2)}_3$ become important.
The origin of the second term may intuitively be interpreted as the reduction
of the first order contribution $I^{(1)}(\Delta_0)$ since quantum 
fluctuations lead to an occupation of the adjacent charge states $n=-1$ and 
$2$ which is only algebraically suppressed, and no more exponentially.
Therefore, the probability of the system to be in state $n=0$ or $1$ (which
is necessary for the first order process) is decreased.
The third term may indicate the appearance of a renormalization of the
excitation energy $\Delta_0$ \cite{SS,KSS1,Mat,Fal-Schoen-Zim}.
Due to this renormalization (which should be proportional to the gap and 
reduce it) the system is effectively ``closer'' to the resonance as the 
original parameters would suggest.
The current would then, in second order, be roughly given by the derivative of 
the first order term times the renormalization.

The behaviour of the system at resonance (and its crossover to the Coulomb 
blockade regime) was also described in Ref.~\cite{SS,KSS1} within the 
resonant tunneling approximation for the two charge state model.
Therefore, the expansion of the resonant tunneling formula up to second order
yields Eqs.~(\ref{current1}) - (\ref{current3}) if we omit all terms 
with $\Delta_1$ and $\Delta_{-1}$.
The integrals, then, become divergent and a cut-off (of the order of the 
charging energy) has to be introduced.
In this Letter, however, we took into account all processes, and, therefore,
no cut-off is needed.

\begin{figure}
\centerline{\psfig{figure=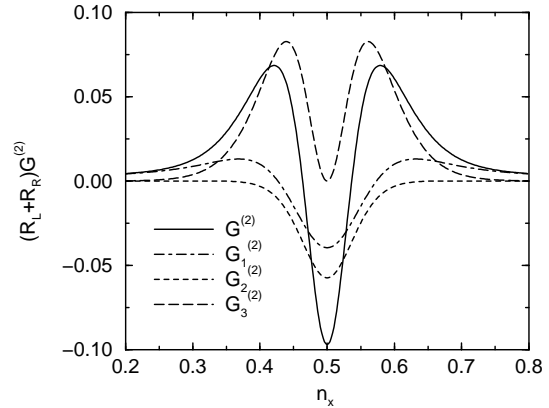,height=6cm}}
\caption{The second-order contribution of the differential conductance 
        $G^{(2)}=\sum_{i=1}^3 G_i^{(2)}$ for $T/E_{\rm C}=0.05$, 
	$\alpha_0=0.04$ and $V=0$.}
\label{fig2}
\end{figure}

\begin{figure}
\centerline{\psfig{figure=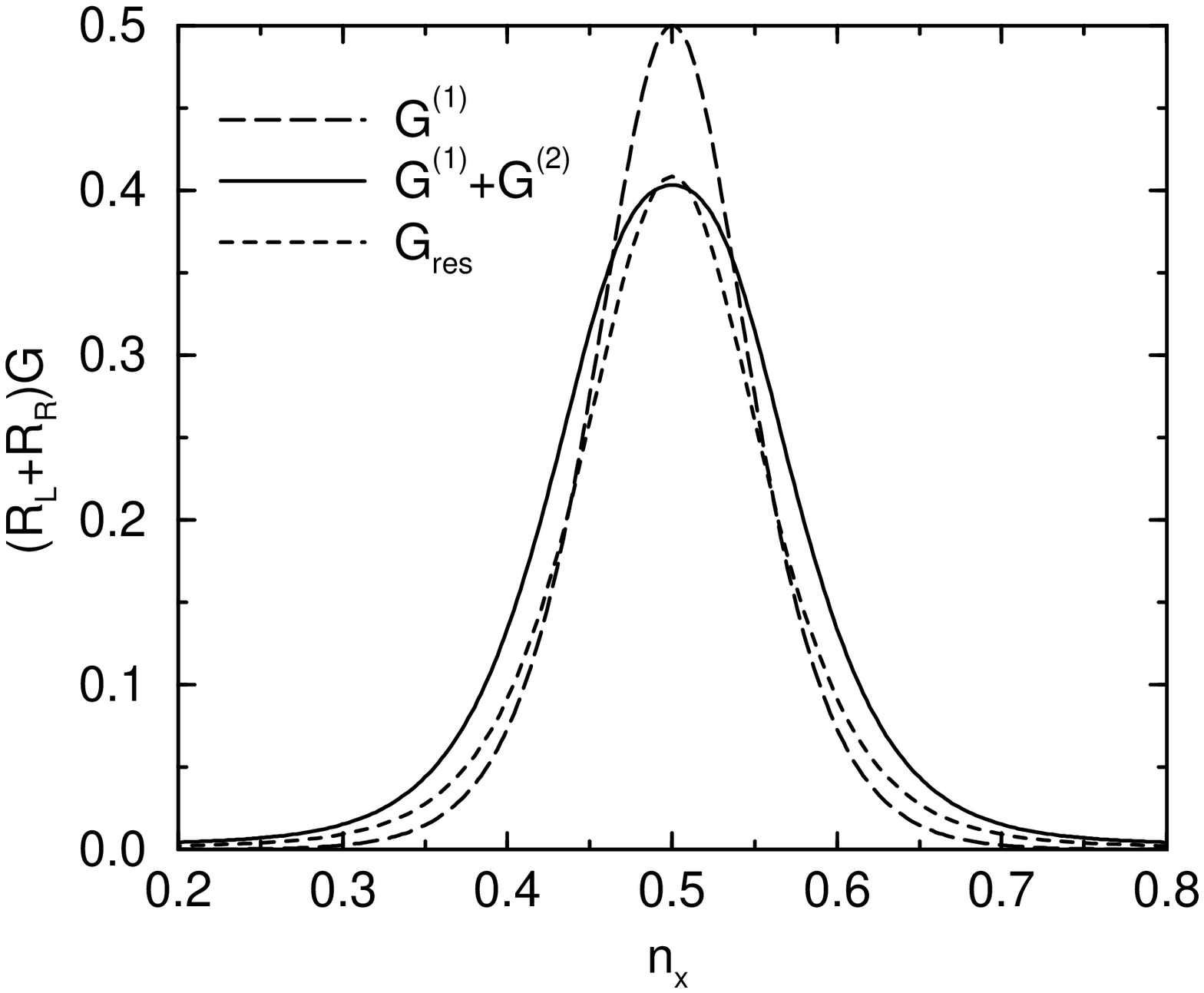,height=6cm}}
\caption{The differential conductance for $T/E_{\rm C}=0.05$, $\alpha_0=0.04$ 
	and $V=0$: sequential tunneling, sequential plus cotunneling, and 
	resonant tunneling approximation.}
\label{fig3}
\end{figure}

\begin{figure}
\centerline{\psfig{figure=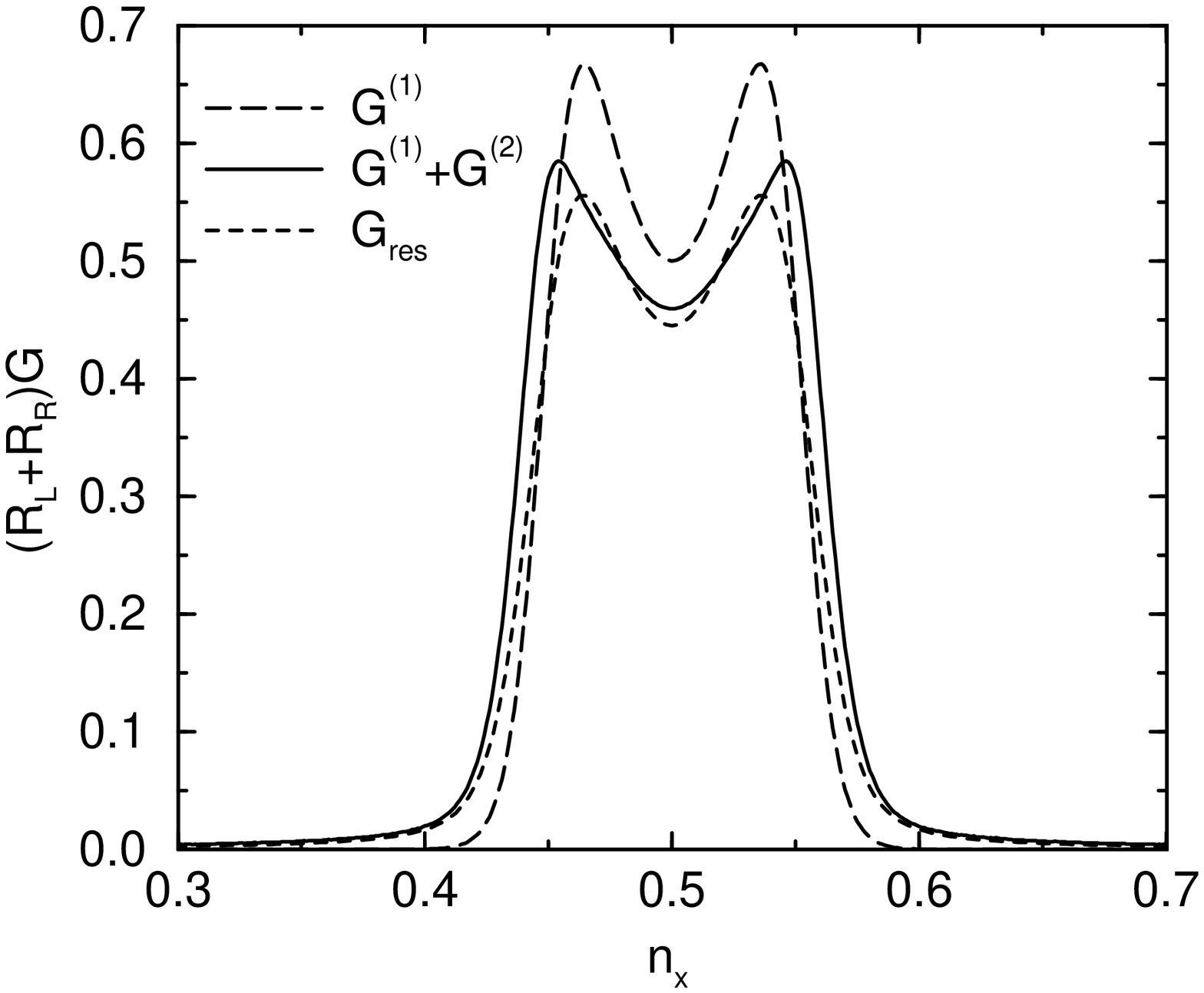,height=6cm}}
\caption{The differential conductance for $T/E_{\rm C}=0.01$, $\alpha_0=0.02$ 
	and $V/E_{\rm C}=0.2$: sequential tunneling, sequential plus 
	cotunneling, and resonant tunneling approximation.}
\label{fig4}
\end{figure}

In Fig.~\ref{fig2} we show the second order contribution
to the linear differential conductance $G=\partial I/ \partial V$.
Here and in the following we choose a symmetric coupling
$\alpha_0^{\rm L}=\alpha_0^{\rm R}$.
In Figs.~\ref{fig3} and \ref{fig4} a comparison of the first order, the sum of
the first and second order, and the resonant tunneling approximation (where
the cut-off is adjusted as $U=E_{\em C}$) is displayed for the linear
and nonlinear regime.
The deviation from the first order result (sequential tunneling) is 
significant and of the order $20\%$. 
The agreement with the resonant tunneling approximation provides a
clear criterium for the choice of the bandwidth cut-off. Furthermore,
and most importantly, it shows 
the existence of a parameter regime where renormalizations of $E_{\rm C}$,
$\alpha_0$, and $\Delta_0$ by higher order charge states can be
neglected although the current deviates significantly from
the classical result. We have checked the significance of third
order terms $\sim \alpha_0^3$ by using the resonant tunneling formula
\cite{SS,KSS1} and exact results for the average charge in third order
at zero temperature \cite{Gra}. For the parameter sets used in the
figures, the deviations to the sum of first and second order terms
were smaller than about $2\%$. Therefore, at not too low temperatures, second 
order perturbation theory is a good approximation even if the
tunneling resistance approaches the quantum resistance.

\begin{figure}
\vspace{-2cm}
\centerline{\psfig{figure=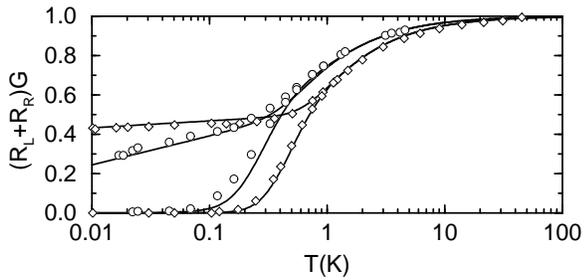,height=7cm}}
\vspace{-1.5cm}
\caption[a]{Maximal and minimal linear conductance for $E_{\rm C}=1.47 K$ and 
 	$\alpha_0=0.015$, and $E_{\rm C}=1 K$ and $\alpha_0=0.063$.
	The dots are experimental data from Ref.~\cite{esteve}.}
\label{fig5}
\end{figure}

In Fig.~\ref{fig5} we present a comparison of our results with recent 
experiments \cite{esteve}. 
The temperature dependence of the Coulomb oscillations were measured for two 
different samples with conductances $\alpha_0 = 0.015$ and $\alpha_0 = 0.063$. 
For $\alpha_0=0.015$, second order perturbation theory is sufficient and the 
results agree perfectly in the whole temperature and gate voltage regime. 
For $\alpha_0=0.063$, third order terms start to become important, but the 
agreement is still reasonable. 
We emphasize, that only the bare values for $\alpha_0$ and $E_{\rm C}$ have 
been used as they were determined unambiguously in the experiment. 
Furthermore, the usage of the full resonant tunneling formula with the bare 
value of the charging energy would lead to a clear deviation from the 
experiment by about $10\%$.
Thus, the inclusion of higher order charge states within second
order perturbation theory, as presented in this Letter, is an
important improvement of the theory. For $\alpha_0=0.063$, the
experiment is at the boarder line where third order terms and
renormalization effects start to become important. We have checked
this by using the resonant tunneling approximation with a renormalized
value of the charging energy as estimated in the experiment. Again,
the agreement is reasonable which indicates that higher order charge
states lead to a renormalization of the charging energy. 

In conclusion we have presented a consistent calculation of the tunneling 
current of the single electron transistor up to second order perturbation 
theory. 
The approach is free of any divergences and provides cut-off independent 
results. 
At resonance we find new terms which are significant for experimentally 
realistic parameters. 
We have found a regime where second order perturbation theory can be used 
without any further renormalization of system parameters. 
A comparison with experiment shows good agreement.

We like to thank D. Esteve, H. Grabert and P. Joyez for stimulating and
useful discussions. Our work was supported by the ``Deutsche
Forschungsgemeinschaft'' as part of ``SFB 195''.

\end{document}